\documentclass[english,preprint]{revtex4}
\usepackage[T1]{fontenc}
\usepackage[latin1]{inputenc}
\usepackage{amsmath}
\usepackage{amssymb}

\makeatletter
\usepackage{amssymb}
\usepackage{babel}

\usepackage{babel}
\makeatother
\begin{document}

\title{Classification of Tripartite Entanglement with one Qubit}

\author{Marcio F. Cornelio}

\email{marcio@fma.if.usp.br}

\author{and A. F. R. de Toledo Piza}

\affiliation{Universidade de São Paulo, Instituto de Física, CP 66318, 05315 São
Paulo, S.P., Brazil}

\begin{abstract}
We present a method to find the decompositions of tripartite entangled
pure states which are smaller than two successive Schmidt decompositions.
The method becomes very simple when one of the subsystems is a qubit.
In this particular case, we get a classification of states according
to their decompositions. Furthermore, we also use this method to classify
the entangled states that can be inter-converted through stochastic
local operations and classical communication (SLOCC). More general
tripartite systems are briefly discussed. 
\end{abstract}
\maketitle

\section{Introduction}

Increasing interest in Quantum Information Theory (QIT) has motivated
the study of general types of quantum entanglement. Although entanglement
is well understood only for systems either of small dimensionality
or involving few parties \cite{bruss}, there is no doubt on the fact
that it has a central importance in QIT. This is mainly due to simple
applications, albeit without any classical analogy, of bipartite entanglement
to quantum communication like teletransportation \cite{teleportation,NielsenChuang},
superdense code \cite{densecode,NielsenChuang} and EPR protocol to
quantum secret key distribution \cite{EPRprotocolQKD,NielsenChuang}.
Furthermore, strong drive for the development of quantum computers
is provided by their apparent intrinsic advantages, as indicated e.g.
by the Shor factoring and Grover search quantum algorithms \cite{NielsenChuang}.
In these algorithms, the coherence of an entangled state of many qubits
is crucial. In this way, a better understanding of general entanglement
is desirable.

To this effect, we notice that the achieved understanding of bipartite
entanglement is mainly based on simple decompositions such as the
Schmidt decomposition \cite{schmidt,NielsenChuang} and also the relative
states decomposition\cite{schro35,schro36,everett}. Moreover, these
decompositions are at the heart of fundamentals papers on the meaning
of entanglement in quantum mechanics since the well know von Neumann
theory of measurement \cite{vonNeumann} and the Einstein-Podolsky-Rosen
incompleteness argument \cite{EPR}. Nowadays, these decompositions
have also been central in the description of the dynamics of the quantum
correlations \cite{NemesPiza1986} and in studies dealing with the
emergence of a `classical' world through the phenomenon of decoherence
\cite{Giulini}. Another feature of these decompositions is that they
are simple, in the sense that entangled state are written as a superposition
of the smallest possible number of product states%
\footnote{In the case of the relative states decomposition, not all decompositions
have this property, but it is always possible fulfill it in infinitely
many ways.%
}. This smallest number is usually referred to as the Schmidt rank
of the entangled state.

For tripartite systems, these decompositions may be applied recursively.
In particular, for three qubit entangled states they give many ways
of representing the entangled state as a superposition of four factorable
states. The use of recursive Schmidt decompositoin is discussed by
Partovi \cite{Partovi}. However, it is already known from the works
of Dür, Vidal and Cirac \cite{dur} and of Acín \emph{et al} \cite{Acin}
that a simpler decomposition with two or three product states exists,
depending on the particular tripartite entangled state one is dealing
with. States that can be written in terms of two product states are
usually said to be of the Greenberger-Horne-Zeilinger (GHZ) type,
or of generalized-GHZ type, and states which require at least three
product components are said to be of the W type.

Furthermore, many applications of entangled states in QIT are related
to the non-local aspect of the quantum entanglement. For these applications,
in which the qubits are spatially separated, it is important to know
in which states $|\psi^{\prime}\rangle$ an entangled state $|\psi\rangle$
can be transformed through local operations. There are many types
of local operations which were extensively discussed by Bennett \emph{et
al} \cite{Bennett}. Here we consider only the class of stochastical
local operations with classical communication (SLOCC). In this case,
if spatially separated observers share an entangled state $|\psi\rangle$,
and are allowed to perform local operations (including measurements
and interacting ancillary systems) on their respective subsystem and
to communicate with each other classically, then they can convert,
with nonvanishing probability of success, the state $|\psi\rangle$
to another state $|\psi^{\prime}\rangle$. If one restricts oneself
to reversible SLOCC, one gets a partition of the set of all states
in classes of different types of entangled states \cite{dur}. In
this way, Dür, Vidal and Cirac \cite{dur} have shown that states
of type GHZ and type W correspond to distinct SLOCC classes. The SLOCC
classification was extend to the four qubit case by Verstraete et
al \cite{Verstraete} and to the case of two qubits and one $n$-level
system by Miyake and Verstraete \cite{Miyake1}. More general aspects
of SLOCC classification were also discussed by Miyake \cite{Miyake2}.

In this work, we start from the observation of Dür, Vidal and Cirac
\cite{dur} that the number of product states in the smallest decomposition
of a state is in general invariant through SLOCC. We then show how
to find these decompositions for tripartite systems. In general, we
show that there are many decompositions which are smaller than that
resulting from two successive Schmidt decompositions, which we call
`sub-Schmidt decompositions' for short. Particularly, for tripartite
systems involving one qubit and local supports with dimensions $n$,
$n$ and 2 (we call local support the subspaces in which the reduced
density matrices of each of the subsystems are non-vanishing), we
get a classification of all decompositions which we use to characterize
all possible SLOCC classes.

The paper is organized as follows. In section \ref{sub:Estados-Emaranhados-em-um-qubit}
we show how to find sub-Schmidt decompositions for entangled states
with local supports $n$, $n$ and 2 and give some examples. In section
\ref{sub:SLOCC} we show how the method for finding such decompositions
can be used to define SLOCC classes. As our demonstrations are all
constructive, our treatment also provides for a way to find SLOCC
protocols to transform entangled states. We close the article with
a discussion on the difficulties of extending the results to more
general tripartite entanglement in section \ref{sub:Casos-mais-gerais}.

\section{\label{sub:Estados-Emaranhados-em-um-qubit}Entangled tripartite
states with one qubit}

In order to get a better understanding of the physical and geometrical
meaning of the algebra which follows, we start with some remarks on
known results obtained by Dür, Vidal and Cirac \cite{dur} for the
entangled three qubit system and on the results obtained by Sanpera
et al. \cite{Sanpera} concerning planes in $C^{2}\otimes C^{2}$
spaces. \textbf{}Sanpera \emph{et al.} \cite{Sanpera} have shown
that a plane generated by two entangled states of a two qubit system
contains either one or two pure states. This result is important because,
when we trace out one of the qubits of the three qubit system, the
local support of the Hilbert space of the two other qubits is at most
bi-dimensional, i.e. a plane. This implies that one does not need
a base for the complete $C^{2}\otimes C^{2}$ space in order to represent
an entangled three qubit state, since a base for its local support
in $C^{2}\otimes C^{2}$ is sufficient. Thus we can always find a
base for its local support with either one or two pure states, and
these alternatives correspond, respectively, to the states of class
W and of class GHZ \cite{dur}. We will follow this line of reasoning
within a different mathematical framework which will make it usefull
also for systems of higher dimensionality.

Consider then a pure state $|\psi\rangle$ in a space $C_{a}^{n}\otimes C_{b}^{n}\otimes C_{c}^{2}$,
where we have labeled the subsystems as $s_{a}$, $s_{b}$ (two n-dimensional
subsystems) and $s_{c}$ (the qubit). Suppose also that our state
$|\psi\rangle$ has local supports with dimensions $n$, $n$ and
2 on the subsystem spaces $C_{a}^{n}$, $C_{b}^{n}$ and $C_{c}^{2}$,
respectively. Note that we need to specify the dimension of the local
supports of the three subsystems, in contrast with bipartite entanglement,
where the local supports always have the same dimension. Of course,
the dimension of any local support can not be greater than the product
of the other two. We will express this situation by saying that \emph{$|\psi\rangle$
is an entangled state of dimensionality $n$ by $n$ by 2} or that
\emph{the entanglement of $\left|\psi\right\rangle $ has dimensionality
$n$ by $n$ by 2 iff their local supports on the subsystems have
dimensions $n$, $n$ and 2 respectively}. We will also denote the
dimensionality of $|\psi\rangle$ by $(n,n,2)$ as a shorthand. Then,
if the tripartite system as a whole is in an entangled state of dimensionality
$\left(n,n,2\right)$, the local support in $s_{ab}=s_{a}+s_{b}$
is a bi-dimensional plane $\mathcal{P}\subset C_{a}^{n}\otimes C_{b}^{n}$.
This plane can be easily found from any relative states decomposition\cite{schro35,schro36,everett}
of $|\psi\rangle$. Explicitly, let $\left\{ |k\rangle\right\} _{k=0,1}$
be a orthonormal base in $C_{c}^{2}$. We can write\begin{equation}
|\psi\rangle=\sum_{k=0,1}c_{k}|r_{k}\rangle|k\rangle,\label{eq:inicial}\end{equation}

\noindent where $|r_{k}\rangle\in\mathcal{P}\subset C_{a}^{n}\otimes C_{b}^{n}$
is the relative state of $|k\rangle\in C_{c}^{2}$ and $|c_{k}|^{2}$
is the probability of finding $s_{c}$ in state $|k\rangle$ or $s_{ab}$
in state $|r_{k}\rangle$. In this way the two states $|r_{k}\rangle$
span the plane $\mathcal{P}$.

Let us now look for the entangled states in $\mathcal{P}$ having
Schmidt rank smaller than $n$ and, so, use them to span $\mathcal{P}$
and write $|\psi\rangle$. Any state $|\phi\rangle$ in $\mathcal{P}$
can be written as a linear combination of the two states $|r_{k}\rangle$,\begin{equation}
\left|\phi\right\rangle =\alpha_{0}\left|r_{0}\right\rangle +\alpha_{1}\left|r_{1}\right\rangle \label{eq:plano}\end{equation}

\noindent where $\alpha_{0}$ and $\alpha_{1}$ are complex coefficients.
In order not to carry unimportant normalization factors, we ignore
the normalization constraint on the coefficients $\alpha_{0}$ and
$\alpha_{1}$. Of course the state can easily be normalized at the
end. Each bipartite entangled state $|\phi\rangle$ can be seen as
a linear mapping of $C_{a}^{n*}$ on $C_{b}^{n}$ (where $C_{a}^{n*}$
is the dual of $C_{a}^{n}$) defined by the partial scalar product
of any $\langle u_{a}|\in C_{a}^{n*}$ with $|\phi\rangle$, $\langle u_{a}|\phi\rangle\in C_{b}^{n}$.
The rank of this linear mapping is the Schmidt rank of the state $|\phi\rangle$.
We are then looking for $\alpha_{0}$ and $\alpha_{1}$ such that
$|\phi\rangle$ has Schmidt rank less than $n$, i. e., we are looking
for $\alpha_{0}$ and $\alpha_{1}$ such that the equation\begin{equation}
\langle u_{a}|(\alpha_{0}|r_{0}\rangle+\alpha_{1}|r_{1}\rangle)=0\label{eq:fundamental}\end{equation}

\noindent has a at least one non-trivial solution $\langle u_{a}|\in C_{a}^{n*}$.

We must emphasize, at least for the moment, that the nature of the
state $|u_{a}\rangle$ is in fact irrelevant, the relevant question
being: which are the $\alpha_{0}$ and $\alpha_{1}$ such that some
non-null $|u_{a}\rangle$ satisfying eq. (\ref{eq:fundamental}) exists?
The state $|u_{a}\rangle$ has an interesting physical meaning, however.
Suppose there is some non-vanishing $|u_{a}\rangle$ for some also
non-vanishing values of $\alpha_{0}$ and $\alpha_{1}$ (since we
have not yet proved that they exist), and suppose further that we
make a measurement on subsystem $s_{a}$ and find it in state $|u_{a}\rangle$.
Then, although the state of $s_{ab}$ is mixed (in particular it can
be seen as a mixture involving $|r_{0}\rangle$ and $|r_{1}\rangle$),
we get also a pure state for $s_{b}$ and, consequently, also for
$s_{c}$ so that the whole system is reduced to a product state. This
happens because the validity of equation (\ref{eq:fundamental}) implies
that $\alpha_{0}\langle u_{a}|r_{0}\rangle=-\alpha_{1}\langle u_{a}|r_{1}\rangle$,
i. e., the vectors $\langle u_{a}|r_{0}\rangle$ and $\langle u_{a}|r_{1}\rangle$
in $C_{b}^{n}$ are linearly dependent. In physical terms, the relative
state for $|u_{a}\rangle$ is the same whether the state of $s_{ab}$
is $|r_{0}\rangle$, $|r_{1}\rangle$ or in fact any state in $\mathcal{P}$
(other than $\alpha_{0}|r_{0}\rangle+\alpha_{1}|r_{1}\rangle$). If
we could find two linearly independent $|u_{a}\rangle$'s for the
same $\alpha_{0}$ and $\alpha_{1}$, we would have a subspace of
$|u_{a}\rangle$'s with the same relative state for all states in
$\mathcal{P}$. This subspace would actually be the null space of
the linear mapping of $C_{a}^{n*}$ on $C_{b}^{n}$ defined by those
particular superpositions of $|r_{0}\rangle$ and $|r_{1}\rangle$
such that $\alpha_{0}\langle u_{a}|r_{0}\rangle=-\alpha_{1}\langle u_{a}|r_{1}\rangle$.
In this way, these particular superpositions would have Schmidt rank
$n$ less the dimension of this subspace of $|u_{a}\rangle$'s. The
existence of states $|u_{a}\rangle$ with this property has already
been observed for three qubits entanglement of dimensionality $(2,2,2)$
by Acín \emph{et al} \cite{Acin}.

In order to find a solution of equation (\ref{eq:fundamental}) we
choose a base $\{|i\rangle\}$ in $C_{a}^{n}$ and a base $\{|j\rangle\}$
in $C_{b}^{n}$, so that we can rewrite it in matrix form as\begin{equation}
\left(\alpha_{0}R_{0}+\alpha_{1}R_{1}\right)u_{a}^{*}=0.\label{eq:superposicao}\end{equation}

\noindent where the matrix $R_{k}$ has components $[R_{k}]_{ij}=\langle ji|r_{k}\rangle$
and the vector $u_{a}^{*}$ has components $u_{a_{i}}^{*}=\langle u_{a}|i\rangle$.
The Schmidt rank of the state $|\phi\rangle=\alpha_{0}|r_{0}\rangle+\alpha|r_{1}\rangle$
is the rank of the matrix $(\alpha_{0}R_{0}+\alpha_{1}R_{1})$. Now
we assume, without loss of generality, that the state $|r_{1}\rangle$
has Schmidt rank $n$. Otherwise we would either have a problem of
dimensionality lower than $(n,n,2)$ or $|r_{1}\rangle$ would already
be a solution of problem. We will show however that the set of solutions
of (\ref{eq:superposicao}) is of null measure in $\mathcal{P}$.
We can rewrite (\ref{eq:superposicao}) so that it looks like an eigenvalue
equation\begin{equation}
(R_{1}^{-1}R_{0}-\lambda)u_{a}^{*}=0\label{eq:DeAutoValores}\end{equation}

\noindent where $\lambda=-\alpha_{1}/\alpha_{0}$. However, we must
keep in mind that we are not solving an eigenvalue problem since $\lambda$
in equation (\ref{eq:DeAutoValores}) depends on the ratio of coefficients
$\alpha_{1}$ and $\alpha_{0}$, which in turn depends on the base
we have chosen in equation (\ref{eq:inicial}) for $\mathcal{P}$.
But the number of distinct states with Schmidt rank smaller than $n$
in $\mathcal{P}$ obviously cannot depend on the base chosen for $\mathcal{P}$,
neither can their respective Schmidt rank.

In this way, we can ask what would change in eq. (\ref{eq:DeAutoValores})
if we would have chosen another base for $\mathcal{P}$ in eq. (\ref{eq:plano}).
Let us call this base $\{|\phi_{k}\rangle\}$, with $|\phi_{0}\rangle=a|r_{0}\rangle+b|r_{1}\rangle$
and $|\phi_{1}\rangle=c|r_{0}\rangle+d|r_{1}\rangle$ where $a$,
$b$, $c$ and $d$ are complex and $(ad-bc)=1$. Thus, analogously
with (\ref{eq:plano}), any state $|\phi\rangle$ in $\mathcal{P}$
can be written as\[
|\phi\rangle=\beta_{0}|\phi_{0}\rangle+\beta_{1}|\phi_{1}\rangle.\]

\noindent With every state $|\phi_{k}\rangle$, we can associate the
matrix $[\Phi_{k}]_{ij}=\langle ji|\phi_{k}\rangle$. We can also
suppose without loss of generality that $\Phi_{1}$ is invertible.
Instead of eq. (\ref{eq:DeAutoValores}), we would thus get\[
(\Phi_{1}^{-1}\Phi_{0}-\mu)u_{a}^{*}=0\]

\noindent where $\mu=-\beta_{1}/\beta_{0}$. What aspects are common
to matrices $R_{1}^{-1}R_{0}$ and $\Phi_{1}^{-1}\Phi_{0}$ and how
are their respective eigenvalues $\lambda_{l}$ and $\mu_{l}$ related?
To answer this question we define the concept of \emph{Jordan family:
we will say that two matrices, $A$ and $B$, are at the same Jordan
family iff for} \emph{each} \emph{eigenvalue $\lambda_{l}$ of $A$
there is an eigenvalue $\mu_{l}$ of $B$ such that the rank of the
matrices $(A-\lambda_{l})^{k}$ and $(B-\mu_{l})^{k}$ are equal for
every positive integer $k$.} This is equivalent to saying that the
Jordan blocks of the matrices $A$ and $B$ in their Jordan canonical
form have the same structure \cite[Section 3.2]{HornJohnson}, although
they can differ in the numerical values of the eigenvalues. The following
theorem asserts that the matrices $R_{1}^{-1}R_{0}$ and $\Phi_{1}^{-1}\Phi_{0}$
are in the same Jordan family.

\textbf{Theorem 1}: \emph{Let $R_{0}$ and $R_{1}$ be two $n$ by
$n$} \emph{matrices, $R_{1}$ invertible, and $\Phi_{0}=aR_{0}+bR_{1}$,
$\Phi_{1}=cR_{0}+dR_{1}$ two linear combinations of $R_{0}$ and
$R_{1}$ such that $\Phi_{1}$ is also invertible and $(ad-bc)=1$.
Then the matrices $R_{1}^{-1}R_{0}$ and $\Phi_{1}^{-1}\Phi_{0}$
belong to the same Jordan family. Moreover, the relation between the
eigenvalues $\lambda_{l}$ of $R_{1}^{-1}R_{0}$ and $\mu_{l}$ of
$\Phi_{1}^{-1}\Phi_{0}$, such that, for all positive integer $k$,}
${\textrm{rank}}(R_{1}^{-1}R_{0}-\lambda_{l})^{k}$=${\textrm{rank}}(\Phi_{1}^{-1}\Phi_{0}-\mu_{l})^{k}$
\emph{is given by}\begin{equation}
\mu_{l}=\frac{a\lambda_{l}+b}{c\lambda_{l}+d}.\label{eq:RelationEigenvalues}\end{equation}

\emph{Proof.}: Let us consider first the case $c=0$, so that $a\neq0$,
$d\neq0$ and straightforward evaluation gives\begin{eqnarray*}
(\Phi_{1}^{-1}\Phi_{0}-\mu_{l})^{k} & = & \left[(dR_{1})^{-1}(aR_{0}+bR_{1})-\frac{a\lambda_{l}+b}{d}\right]^{k}\\
 & = & \frac{a^{k}}{d^{k}}(R_{1}^{-1}R_{0}-\lambda_{l})^{k},\end{eqnarray*}

\noindent from which the desired result follows. In case $c\neq0$,
we must have $c\lambda_{l}+d\neq0$, since $c\lambda_{l}+d$ is an
eigenvalue of $\Phi_{1}^{-1}$ which is invertible. In this case we
first notice that\begin{eqnarray*}
\Phi_{1}^{-1}\Phi_{0} & = & (cR_{0}+dR_{1})^{-1}(aR_{0}+bR_{1})\\
 & = & [R_{1}(cR_{1}^{-1}R_{0}+d)]^{-1}R_{1}(aR_{1}^{-1}R_{0}+b)\\
 & = & (cR_{1}^{-1}R_{0}+d)^{-1}(aR_{1}^{-1}R_{0}+b).\end{eqnarray*}

\noindent Dividing the polynomial $az+b$ by $cz+d$ we get complex
$\gamma$ and $\delta$ such that\[
az+b=\gamma(cz+d)+\delta,\]

\noindent for any complex $z$. Thus, it is clear that $aR_{1}^{-1}R_{0}+b=\gamma(cR_{1}^{-1}R_{0}+d)+\delta$.
Thus, we have\begin{eqnarray*}
\Phi_{1}^{-1}\Phi_{0} & = & (cR_{1}^{-1}R_{0}+d)^{-1}[\gamma(cR_{1}^{-1}R_{0}+d)+\delta]\\
 & = & \gamma+\delta(cR_{1}^{-1}R_{0}+d)^{-1}\end{eqnarray*}

\noindent and\begin{eqnarray*}
(\Phi_{1}^{-1}\Phi_{0}-\mu_{l})^{k} & = & \left(\gamma+\delta(cR_{1}^{-1}R_{0}+d)^{-1}-\gamma-\frac{\delta}{c\lambda_{l}+d}\right)^{k}\\
 & = & \delta^{k}c^{k}(c\lambda_{l}+d)^{-k}(cR_{1}^{-1}R_{0}+d)^{-k}(\lambda_{l}-R_{1}^{-1}R_{0})^{k}.\end{eqnarray*}

\noindent As $\Phi_{0}$ and $\Phi_{1}$ are linearly independent,
$\delta\neq0$. Moreover, as $\Phi_{1}$ is invertible, $c\lambda_{l}+d\neq0$,
since its an eigenvalue of $\Phi_{1}$, and $cR_{1}^{-1}R_{0}+d$
is also invertible. Then we have that $\textrm{{rank}}(\Phi_{1}^{-1}\Phi_{0}-\mu_{l})^{k}$=
${\textrm{rank}}(\lambda_{l}-R_{1}^{-1}R_{0})^{k}$ as desired. $\Box$

Therefore, for each eigenvalue $\lambda_{l}$ found using a base $\{|r_{0}\rangle,|r_{1}\rangle\}$,
the use of a different base $\{|\phi_{0}\rangle,|\phi_{1}\rangle\}$,
would also give a corresponding eigenvalue $\mu_{l}$. Moreover, the
rank of the matrix $(R_{1}^{-1}R_{0}-\lambda_{l})^{k}$ is equal to
the rank of $(\Phi_{1}^{-1}\Phi_{0}-\mu_{l})^{k}$. This rank for
$k=1$ (first rank for short) is simply the Schmidt rank of the state
$|\phi_{l}\rangle=\alpha_{0l}|r_{0}\rangle+\alpha_{1l}|r_{l}\rangle=\beta_{0l}|\phi_{0}\rangle+\beta_{1l}|\phi_{1}\rangle$,
where $\alpha_{0l}$ and $\alpha_{1l}$ are such that $\lambda_{l}=-\alpha_{1l}/\alpha_{0l}$
and $\beta_{0l}$ and $\beta_{1l}$ are such that $\mu_{l}=-\beta_{1l}/\beta_{0l}$.
Thus, the first rank can be understood as a property of the state
$|\phi_{l}\rangle$ alone, since it will not change if the same state
$|\phi_{l}\rangle$ is also found in another plane. The same is not
true for the higher ranks ($k\geq2$), which can be different for
the same $|\phi_{l}\rangle$ in different planes. In this way, we
must understand these higher ranks as invariant properties of the
state $|\phi_{l}\rangle$ inside the plane $\mathcal{P}$ and also
as invariant properties of the whole tripartite entanglement state
$\left|\psi\right\rangle $ from which $\mathcal{P}$ is obtained.
The distinction of these higher ranks is important for $n\geq4$.
We write these states explicitly for entangled states of dimensionality
$(4,4,2)$ in example 3.

If subsystems $s_{a}$ and $s_{b}$ are interchanged the result is
equivalent. In this case the states $|r_{0}\rangle$ and $|r_{1}\rangle$
in eq. (\ref{eq:fundamental}) must be understood as linear mappings
from $C_{b}^{n*}$ in $C_{a}^{n}$ and, instead of the matrix $R_{0}$
and $R_{1}$ in eq. (\ref{eq:superposicao}), we will get their respective
transposes $R_{0}^{T}$ and $R_{1}^{T}$. Thus, in place of the matrix
$R_{1}^{-1}R_{0}$ in eq. (\ref{eq:DeAutoValores}), we will get the
matrix $(R_{0}R_{1}^{-1})^{T}$ which is similar to it.

We can now use the solutions of equation (\ref{eq:DeAutoValores})
to find states in $\mathcal{P}$ with Schmidt rank smaller than $n$
and then rewrite the state $|\psi\rangle$ in terms of a smaller number
of product states. More explicitly, suppose, we have found two solutions
and then write its respective normalized bipartite states $|\phi_{1}\rangle$
and $|\phi_{2}\rangle$ in $\mathcal{P}$ with Schmidt rank smaller
than $n$. Using the base $\{|\phi_{1}\rangle,|\phi_{2}\rangle\}$
to span $\mathcal{P}$, we have\begin{equation}
|\psi\rangle=|\phi_{1}\rangle|c_{1}\rangle+|\phi_{2}\rangle|c_{2}\rangle,\label{eq:decMinima}\end{equation}

\noindent where $|c_{1}\rangle$ and $|c_{2}\rangle$ are appropriate
non-normalized states in $C_{c}^{2}$ , given by\begin{eqnarray*}
|c_{1}\rangle & = & \sum_{k}c_{k}|k\rangle(g_{11}\langle\phi_{1}|r_{k}\rangle+g_{12}\langle\phi_{2}|r_{k}\rangle)\\
|c_{2}\rangle & = & \sum_{k}c_{k}|k\rangle(g_{21}\langle\phi_{1}|r_{k}\rangle+g_{22}\langle\phi_{2}|r_{k}\rangle),\end{eqnarray*}

\noindent The metric coefficients $g_{ij}$ appear from the fact that
$|\phi_{1}\rangle$ and $|\phi_{2}\rangle$ are in general non-orthogonal
states. We must observe that $|c_{1}\rangle$ and $|c_{2}\rangle$
are in general also non-orthogonal. From (\ref{eq:decMinima}), it
is easy to see that $|\psi\rangle$ may be written in terms of a number
of products given by the sum of the Schmidt ranks of $|\phi_{1}\rangle$
and $|\phi_{2}\rangle$. Equation (\ref{eq:DeAutoValores}) may have
from one to $n$ distinct eigenvalues. In the case of a single eigenvalue,
we have just one state with Schmidt rank smaller than $n$ and will
have to choose another state in $\mathcal{P}$ with Schmidt rank $n$.
When we find $m\geq2$ solutions of equation (\ref{eq:DeAutoValores}),
we have $\left(\begin{array}{c}
m\\
2\end{array}\right)$ distinct combinations of $\left|\phi_{1}\right\rangle $ and $\left|\phi_{2}\right\rangle $
to write $|\psi\rangle$ in (\ref{eq:decMinima}). Of course, it is
always possible to choose a state with Schmidt rank smaller than $n$
and another with Schmidt rank $n$ and each $\left|\phi_{k}\right\rangle $
has also infinitaly many bipartite decomposition. Then there will
be always infinitaly many sub-Schmidt decompositions of $\left|\psi\right\rangle $
in (\ref{eq:decMinima}).

Therefore, for each Jordan family to which the matrix $R_{1}^{-1}R_{0}$
may belong we can associate a family of entangled states $|\psi\rangle$.
These entangled states will be all of dimensionality $(n,n,2)$, except
for the family of matrices proportional to the identity matrix which
implies that matrices $R_{0}$ and $R_{1}$ are proportional and,
therefore, that the qubit is not entangled with the other two $n$-dimensional
subsystems, i. e., in this case, we have a ordinary bipartite entanglement
with Schmidt rank $n$ of the two $n$-dimensional subsystems.

We will see in the following section that states which belong to distinct
families in fact belong to distinct SLOCC classes. Before discussing
the relation between the Jordan families and SLOCC classification,
it is convenient to discuss some examples of Jordan families and their
respective sub-Schmidt decomposition of their corresponding entangled
states.

\subsection{Example 1: Three qubits.}

The properties of three qubit entangled states are well know \cite{dur,bruss,Acin}.
Here we reproduce known results for this case in terms of the procedure
described above as an example of its use. A state $|\psi\rangle$
with entanglement of dimensionality $(2,2,2)$ can be identified with
one of the following two Jordan families\[
\textrm{(a):}\;\,\left(\begin{array}{cc}
\lambda_{1} & 1\\
0 & \lambda_{1}\end{array}\right)\hspace{1.5cm}\textrm{(b):}\;\,\left(\begin{array}{cc}
\lambda_{1} & 0\\
0 & \lambda_{2}\end{array}\right)\]

\noindent where $\lambda_{1}\neq\lambda_{2}$. In case (a), there
is only one state\[
|\phi_{\lambda_{1}}\rangle=\alpha_{\lambda_{1}0}|r_{0}\rangle+\alpha_{\lambda_{1}1}|r_{1}\rangle,\]

\noindent where $\lambda_{1}=-\alpha_{\lambda_{1}1}/\alpha_{\lambda_{1}0}$,
with Schmidt rank 1 in $\mathcal{P}$, i.e., $|\phi_{\lambda_{1}}\rangle$
is the only unentangled state in $\mathcal{P}$. Then, if we want
to span $\mathcal{P}$, we have to choose another state $|\phi_{2}\rangle\in\mathcal{P}$
with Schmidt rank 2 in $\mathcal{P}$. From (\ref{eq:decMinima}),
it follows that $|\psi\rangle$ can be written as a superposition
of three product states. This means that $|\psi\rangle$ is in class
W \cite{dur}, since it can be converted through some SLOCC to the
state\[
|W\rangle=\frac{1}{\sqrt{3}}(|001\rangle+|010\rangle+|100\rangle).\]

\noindent In case (b), on the other hand, we have two unentangled
states in $\mathcal{P}$, one for each $\lambda_{l}$, given by\[
|\phi_{\lambda_{l}}\rangle=\alpha_{\lambda_{l}0}|r_{0}\rangle+\alpha_{\lambda_{l}1}|r_{1}\rangle,\]

\noindent where $\lambda_{l}=-\alpha_{\lambda_{l}1}/\alpha_{\lambda_{l}0}$.
Then we can write $|\psi\rangle$ as a superposition of two product
states, meaning that $|\psi\rangle$ belongs to class GHZ, since it
can be converted through some SLOCC to the state \cite{dur}\[
|GHZ\rangle=\frac{1}{\sqrt{2}}(|000\rangle+|111\rangle).\]

\noindent We observe that our method provides for a way to decide
whether a given state is in class W or in class GHZ which is alternate
to that proposed in \cite{dur}. The dimensionality of entanglement
can easily be obtained from the determinant of the reduced density
matrices of subsystems. Once we have verified that a state $\left|\psi\right\rangle $
involves an entanglement of dimensionality $\left(2,2,2\right)$,
we just have to verify whether equation (\ref{eq:DeAutoValores})
has one or two solutions. With little further calculation we can also
get the sub-Schmidt decompositions. We will also see in section \ref{sub:SLOCC}
that we can constructively determine the SLOCC which transforms the
considered state into the state $|W\rangle$ or $|GHZ\rangle$. All
other states with entanglement dimensionality smaller than $\left(2,2,2\right)$
(smaller meaning that at least one of local supports has smaller dimensionality
and none has higher) show ordinary bipartite entanglement or are completely
unentangled states.

\subsection{Example 2: one qubit and two three level systems.}

In this example, we show new families of entangled states which are
simple to write down and which provide insight for more general systems
with higher entanglement dimensionality. Let $|\psi\rangle$ be an
entangled state of dimensionality $(3,3,2)$. Then $|\psi\rangle$
must be in one of the following five Jordan families:\begin{eqnarray*}
 &  & \textrm{(a):}\;\,\left(\begin{array}{ccc}
\lambda_{1} & 1 & 0\\
0 & \lambda_{1} & 1\\
0 & 0 & \lambda_{1}\end{array}\right)\hspace{1cm}\textrm{(b):}\,\;\left(\begin{array}{ccc}
\lambda_{1} & 0 & 0\\
0 & \lambda_{1} & 1\\
0 & 0 & \lambda_{1}\end{array}\right)\\
 &  & \textrm{(c):}\,\;\left(\begin{array}{ccc}
\lambda_{1} & 0 & 0\\
0 & \lambda_{2} & 1\\
0 & 0 & \lambda_{2}\end{array}\right)\hspace{1cm}\textrm{(d):}\;\,\left(\begin{array}{ccc}
\lambda_{1} & 0 & 0\\
0 & \lambda_{2} & 0\\
0 & 0 & \lambda_{2}\end{array}\right)\\
 &  & \hspace{2cm}\textrm{(e):}\;\,\left(\begin{array}{ccc}
\lambda_{1} & 0 & 0\\
0 & \lambda_{2} & 0\\
0 & 0 & \lambda_{3}\end{array}\right)\end{eqnarray*}

\noindent where $\lambda_{l}\neq\lambda_{l^{\prime}}$ for $l\neq l^{\prime}$.
For each one of this families:

\textbf{(a):} There is only one $|\phi_{\lambda_{1}}\rangle$ with
Schmidt rank 2 in $\mathcal{P}$. Then, to span $\mathcal{P}$, we
need to choose another $|\phi_{2}\rangle\in\mathcal{P}$ with Schmidt
rank 3. From (\ref{eq:decMinima}), we get that $|\psi\rangle$ can
be written as a superposition of five product states. We choose as
the characteristic example of this family the state\[
|\psi_{a}\rangle=\frac{1}{\sqrt{5}}[(|10\rangle+|21\rangle)|0\rangle+(|00\rangle+|11\rangle+|22\rangle)|1\rangle].\]

\textbf{(b):} There is only one state $|\phi_{\lambda_{1}}\rangle$
with Schmidt rank 1 in $\mathcal{P}$. Then we need to choose another
$|\phi_{2}\rangle\in\mathcal{P}$ with Schmidt rank 3 to write $|\psi\rangle$
as a superposition of four product states. We choose as the characteristic
example of this family the state\[
|\psi_{b}\rangle=\frac{1}{2}[|21\rangle\left|0\right\rangle +(|00\rangle+|11\rangle+|22\rangle)|1\rangle].\]

\textbf{(c):} There are two states, $|\phi_{\lambda_{1}}\rangle$
and $|\phi_{\lambda_{2}}\rangle$ with Schmidt rank 2 in $\mathcal{P}$.
Then we can use them to span $\mathcal{P}$ and, using (\ref{eq:decMinima}),
write $|\psi\rangle$ as a superposition of four product states. We
choose as the characteristic example of this family the state\[
|\psi_{c}\rangle=\frac{1}{2}[(|00\rangle+|21\rangle)|0\rangle+(|11\rangle+|22\rangle)|1\rangle].\]

\textbf{(d):} There is one state $|\phi_{\lambda_{1}}\rangle$ with
Schmidt rank 1 and also one state $|\phi_{\lambda_{1}}\rangle$ with
Schmidt rank 2 in $\mathcal{P}$. Then we can use them to write $|\psi\rangle$
as a superposition of three product states. We choose as the characteristic
example of this family the state\[
|\psi_{d}\rangle=\frac{1}{\sqrt{3}}[|00\rangle\left|0\right\rangle +(|11\rangle+|22\rangle)|1\rangle].\]

\textbf{(e):} There are three states $|\phi_{\lambda_{l}}\rangle$
with Schmidt rank 2 in $\mathcal{P}$. As we need only two to span
$\mathcal{P}$, we have three ways in (\ref{eq:decMinima}) to write
$|\psi\rangle$ as a superposition of four product states. We choose
as the characteristic example of this family the state\[
|\psi_{e}\rangle=\frac{1}{2}[(|00\rangle+|11\rangle)|0\rangle+(|11\rangle+|22\rangle)|1\rangle].\]

Therefore, an entangled state $|\psi\rangle$ of dimensionality $(3,3,2)$
can be classified in five distinct Jordan families which correspond
to five distinct ways of sub-Schmidt decomposing it in terms of 3,
4 or 5 product states. Moreover, we see that there are three families
with a sub-Schmidt decomposition of 4 product states. We see therefore
that is not just the number of product states that distinguishes entangled
states, but also the nature of the decomposition and the number of
distinct decompositions (compare e.g. cases (b), (c) and (d), which
involve four product states). As will be shown in section \ref{sub:SLOCC},
each one of these families corresponds to a distinct SLOCC class.

\subsection{Example 3: one qubit and two four level subsystems.}

We will not list explicitly all families for the entangled states
of dimensionality $(4,4,2)$. There are in all thirteen families,
and we will limit ourselves to discuss those which emphasize some
aspects that did not arise in connection with example 2.

We start with the following situation. Suppose that Alice, Bob and
Carol share three qubits in a state $|GHZ\rangle$ or $|W\rangle$
and that Alice and Bob also share two qubits in the Bell state $|\phi^{+}\rangle=\frac{1}{\sqrt{2}}(|00\rangle+|11\rangle)$.
Then we can consider Alice's and Bob's two qubits as our four level
subsystems and the state of all the five qubits is an entangled state
of dimensionality $(4,4,2)$. The sub-Schmidt decompositions of these
two states are most easily obtained directly from the evaluation of
the tensor products $|GHZ\rangle\otimes|\phi^{+}\rangle$ and $|W\rangle\otimes|\phi^{+}\rangle$,
i.e.,\[
|GHZ\rangle\otimes|\phi^{+}\rangle=\frac{1}{2}[(|00,00\rangle+|01,01\rangle)|0\rangle+(|10,10\rangle+|11,11\rangle)|1\rangle]\]

\noindent and\[
|W\rangle\otimes|\phi^{+}\rangle=\frac{1}{\sqrt{6}}[(|00,00\rangle+|01,01\rangle)|1\rangle+(|00,10\rangle+|01,11\rangle+|10,00\rangle+|11,01\rangle)|0\rangle].\]

\noindent Our procedure further reveals that these sub-Schmidt decompositions
are \textbf{}the smallest ones and that these states belong respectively
to the following Jordan families\[
(a):\;\, A=\,\left(\begin{array}{cccc}
\lambda_{1} & 0 & 0 & 0\\
0 & \lambda_{1} & 0 & 0\\
0 & 0 & \lambda_{2} & 0\\
0 & 0 & 0 & \lambda_{2}\end{array}\right)\,\,\,\,\,\,\,\,\textrm{and}\,\,\,\,\,\,\,\,(b):\;\, B=\,\left(\begin{array}{cccc}
\lambda_{1} & 1 & 0 & 0\\
0 & \lambda_{1} & 0 & 0\\
0 & 0 & \lambda_{1} & 1\\
0 & 0 & 0 & \lambda_{1}\end{array}\right).\]

\noindent Note that the Jordan family corresponding to $B$ differs
from that corresponding to\[
(c):\;\, C=\,\left(\begin{array}{cccc}
\lambda_{1} & 0 & 0 & 0\\
0 & \lambda_{1} & 1 & 0\\
0 & 0 & \lambda_{1} & 1\\
0 & 0 & 0 & \lambda_{1}\end{array}\right)\]

\noindent only in that the ranks of $(B-\lambda_{1})^{k}$ and $(C-\lambda_{1})^{k}$
differ for $k=2$. In this way, the local support planes in $C_{b}^{4}\otimes C_{c}^{4}$,
$\mathcal{P}_{b}$ and $\mathcal{P}_{c}$, of any state belonging
to one of the families $(b)$ or $(c)$ will have only one state $|\phi_{\lambda_{1}}\rangle$
with Schmidt rank 2 and all other states in $\mathcal{P}_{b}$ and
$\mathcal{P}_{c}$ will have Schmidt rank 4. Thus any state in families
$(b)$ or $(c)$ will have a smallest decomposition with six products
states. An example of a state in this family $(c)$ is\begin{eqnarray*}
\left|\psi_{c}\right\rangle  & = & \frac{1}{\sqrt{6}}[(|10,01\rangle+|11,10\rangle)|1\rangle+(|00,00\rangle+|01,01\rangle+|10,10\rangle+|11,11\rangle)|0\rangle].\end{eqnarray*}

\noindent Other Jordan families that differ in the higher-$k$ ranks
are\[
(d):\;\, D=\left(\begin{array}{cccc}
\lambda_{1} & 0 & 0 & 0\\
0 & \lambda_{2} & 1 & 0\\
0 & 0 & \lambda_{2} & 1\\
0 & 0 & 0 & \lambda_{2}\end{array}\right)\,\,\,\,\,\,\,\,\
\textrm{and}\,\,\,\,\,\,\,\,(e):\;\, E=\left(\begin{array}{cccc}
\lambda_{1} & 1 & 0 & 0\\
0 & \lambda_{1} & 0 & 0\\
0 & 0 & \lambda_{2} & 1\\
0 & 0 & 0 & \lambda_{2}\end{array}\right).\]

\noindent The ranks of $(D-\lambda_{1})^{k}$ and of $(E-\lambda_{1})^{k}$
differ for $k\geq2$, while the ranks of $(D-\lambda_{2})^{k}$ and
of $(E-\lambda_{2})^{k}$ differ for $k\geq3$. Their respective sub-Schmidt
decompositions will also involve six product states. Examples of states
of these classes are\[
|\psi_{d}\rangle=\frac{1}{\sqrt{6}}[(|11\rangle+|22\rangle+|33\rangle)|0\rangle+(|00\rangle+|21\rangle+|32\rangle)|1\rangle]\]

\noindent and\[
|\psi_{e}\rangle=\frac{1}{\sqrt{6}}[(|10\rangle+|22\rangle+|33\rangle)|0\rangle+(|00\rangle+|11\rangle+|32\rangle)|1\rangle]\]

\noindent belonging to Jordan families $(d)$ and $(e)$, respectively.

Two other families which are distinguished in higher-$k$ ranks, $k\geq2$,
for either of the two eigenvalues are\[
(f):\,\left(\begin{array}{cccc}
\lambda_{1} & 0 & 0 & 0\\
0 & \lambda_{2} & 0 & 0\\
0 & 0 & \lambda_{2} & 1\\
0 & 0 & 0 & \lambda_{2}\end{array}\right)\,\,\,\,\,\,\,\,\textrm{and}\,\,\,\,\,\,\,\,(g):\,\left(\begin{array}{cccc}
\lambda_{1} & 1 & 0 & 0\\
0 & \lambda_{1} & 0 & 0\\
0 & 0 & \lambda_{2} & 0\\
0 & 0 & 0 & \lambda_{2}\end{array}\right).\]

\noindent Examples of states in these families are\[
|\psi_{f}\rangle=\frac{1}{\sqrt{5}}[(|11\rangle+|22\rangle+|33\rangle)|0\rangle+(|00\rangle+|23\rangle)|1\rangle]\]

\noindent and\[
|\psi_{g}\rangle=\frac{1}{\sqrt{5}}[(|22\rangle+|33\rangle+|10\rangle)|0\rangle+(|00\rangle+|11\rangle)|1\rangle].\]

Another interesting family is\[
\left(h\right)\,\left(\begin{array}{cccc}
\lambda_{1} & 0 & 0 & 0\\
0 & \lambda_{2} & 0 & 0\\
0 & 0 & \lambda_{3} & 0\\
0 & 0 & 0 & \lambda_{4}\end{array}\right),\]

\noindent which is the only one at this entanglement dimensionality
that needs to be subdivided into an infinity of SLOCC classes as will
be seen in section \ref{sub:SLOCC}. The existence of infinitely many
SLOCC classes for entanglements of higher dimensionality was already
noted by Dür \emph{et al} \cite{dur} using a counting parameter argument.
We can write as an example of a state in this family the state\[
|\psi_{h}\rangle=\frac{1}{\sqrt{4+2\left|a\right|^{2}}}[(|11\rangle+a|22\rangle+|33\rangle)|0\rangle+(|00\rangle+a|11\rangle+|22\rangle)|1\rangle],\]

\noindent where $a\neq0$, so that the associated component does not
vanish, and $a\neq1$, which reflects the fact that all components
cannot be made simultaneously equal. This state has also five more
sub-Schmidt decompositions.

\section{\label{sub:SLOCC}Sub-Schmidt Decompositions and SLOCC}

In this section we discuss the relation between the representation
developed in section \ref{sub:Estados-Emaranhados-em-um-qubit} and
transformation of entangled states through SLOCC protocols. We will
start studying the relation between the Jordan canonical forms of
the matrix $R_{1}^{-1}R_{0}$ in eq. (\ref{eq:DeAutoValores}) derived
from two states that are interconvertible through some SLOCC. After
this, with further analysis of the relation between the eigenvalues
given by eq. (\ref{eq:RelationEigenvalues}), we give a simple criterion
to verify whether two given entangled states of dimensionality $(n,n,2)$
are related by SLOCC.

In order to determine whether a pure state $|\psi\rangle$ can be
transformed into a state $|\psi^{\prime}\rangle$ through SLOCC we
can use the following theorem given in \cite{dur}: a pure state $|\psi\rangle$
can be transformed into a pure state $|\psi^{\prime}\rangle$ through
a SLOCC iff a local linear operator $A\otimes B\otimes C$ exists
such that

\begin{equation}
|\psi^{\prime}\rangle=A\otimes B\otimes C|\psi\rangle\label{eq:LO-SLOCC}\end{equation}

\noindent where $A$, $B$ and $C$ are linear operators in $C_{a}^{n}$,
$C_{b}^{n}$ and $C_{c}^{2}$ respectively%
\footnote{In fact, we must have $A^{\dagger}A$, $B^{\dagger}B$ and $C^{\dagger}C\leq1$,
since $A$ must come from the POVM defined by operators $\sqrt{w_{A}}A$
and $\sqrt{1_{A}-w_{A}A^{\dagger}A}$ performed by Alice, where $w_{A}$
is some positive weight and $1_{A}$ is the identity in $C_{a}^{n}$,
and similarly for $B$ and C. Note that $w_{A}\leq\frac{1}{\lambda_{max}}$
where $\lambda_{max}$ is the greatest eigenvalue of $A^{\dagger}A$.
However, we do not normalize the operators to simplify the calculation.
The final state can always be easily normalized\cite{dur}.%
}. If we consider only invertible local linear operators in (\ref{eq:LO-SLOCC})
we get an equivalence relation between $|\psi\rangle$ and $|\psi^{\prime}\rangle$
which corresponds to the same equivalence relation defined by invertible
SLOCC.

Let us consider what happens when using this result on the decompositions
developed in the preceding section. Suppose that relation (\ref{eq:LO-SLOCC})
is satisfied for some entangled states $|\psi\rangle$ and $|\psi^{\prime}\rangle$
of dimensionality $(n,n,2)$ and some invertible linear operators
$A$, $B$ and $C$. Then, writing $|\psi\rangle$ as in (\ref{eq:inicial})
and inserting in (\ref{eq:LO-SLOCC}), we have

\begin{equation}
|\psi^{\prime}\rangle=\sum_{k=0,1}A\otimes B|r_{k}\rangle C(c_{k}|k\rangle)=\sum_{k=0,1}|\phi_{k}^{\prime}\rangle|c_{k}^{\prime}\rangle,\label{PsiLinhaSLOCCPsi}\end{equation}

\noindent where $|c_{k}^{\prime}\rangle=C(c_{k}|k\rangle)$ and $|\phi_{k}^{\prime}\rangle=A\otimes B|r_{k}\rangle$.
Note that the states $|c_{k}^{\prime}\rangle$ are in general non-orthogonal,
and that therefore the states $|\phi_{k}^{\prime}\rangle$ are not
in general the corresponding relative states of $|c_{k}^{\prime}\rangle$
for $|\psi^{\prime}\rangle$. Since we did not normalize the operators
$A$, $B$ and $C$, the states $|c_{k}^{\prime}\rangle$, $|\phi_{k}^{\prime}\rangle$
and consequently $|\psi^{\prime}\rangle$ are not normalized. We observe
that the invertible linear operator $C$ can transform the states
$c_{k}|k\rangle$ into any two distinct states $|c_{k}^{\prime}\rangle$.
The operator $A\otimes B$ is obviously not so general in $C_{a}^{n}\otimes C_{b}^{n}$.
A well know fact is that it preserves the Schmidt rank of any state
\cite{dur}. In order to understand this, we will study the relation
between the local supports $\mathcal{P}$ and $\mathcal{P}^{\prime}$
of $|\psi\rangle$ and $|\psi^{\prime}\rangle$, respectively, in
$C_{a}^{n}\otimes C_{b}^{n}$. It is clear that, for any $|\phi\rangle\in\mathcal{P}$,
there is a unique $|\phi^{\prime}\rangle\in\mathcal{P}^{\prime}$
such that $|\phi^{\prime}\rangle=A\otimes B|\phi\rangle$. Particularly,
for each one of the states $|r_{k}\rangle$, we have $|\phi_{k}^{\prime}\rangle=A\otimes B|r_{k}\rangle$.
Writing the operators $A$ and $B$ in the local bases of their respective
subsystems, i. e., $\{|i\rangle\}$ for $A$ in $C_{a}^{n}$ and $\{|j\rangle\}$
for $B$ in $C_{b}^{n}$, and the matrices $[R_{k}]_{ij}=\langle ji|r_{k}\rangle$
and $[\Phi_{k}^{\prime}]_{ij}=\langle ji|\phi_{k}^{\prime}\rangle$
as we did in (\ref{eq:superposicao}), we get\[
\Phi_{k}^{\prime}=BR_{k}A^{T}\]
where the matrices $R_{k}$ and $\Phi_{k}^{\prime}$ have the compenents
$[R_{k}]_{ij}=\langle ji|r_{k}\rangle$ and $[\Phi_{k}^{\prime}]_{ij}=\langle ji|\phi_{k}^{\prime}\rangle$,
respectivaly, as in section \ref{sub:Estados-Emaranhados-em-um-qubit}.
Thus, if we use the base $\left\{ |\phi_{k}^{\prime}\rangle\right\} $
to evaluate (\ref{eq:DeAutoValores}) for the state $|\psi^{\prime}\rangle$,
we will get\begin{equation}
\Phi_{1}^{\prime-1}\Phi_{0}^{\prime}=A^{T^{-1}}R_{1}^{-1}R_{0}A^{T}.\label{eq:FiSimilarR}\end{equation}
Then, we see that the matrices $R_{1}^{-1}R_{0}$ and $\Phi_{1}^{\prime-1}\Phi_{0}^{\prime}$
are similar, i. e., they have the same Jordan canonical form. Note
that $\Phi_{1}^{\prime}$ is invertible iff $R_{1}$ is invertible.

The fact that the operator $B$ no longer appears in eq. (\ref{eq:FiSimilarR})
does not imply that it is unimportant, since the plane $\mathcal{P}^{\prime}$
generated by the base $\{ A\otimes B|r_{0}\rangle,A\otimes B|r_{1}\rangle\}$
is obviously distinct from the plane $\mathcal{P}^{\prime\prime}$
generated by the base $\{ A\otimes I_{b}|r_{0}\rangle,A\otimes I_{b}|r_{1}\rangle\}$,
although they result in the same matrix $A^{T^{-1}}R_{1}^{-1}R_{0}A^{T}$.
Moreover, if we interchange the roles of subsystem $s_{a}$ and $s_{b}$,
we find that $A$ is the operator which that no longer appears between
matrices $R_{1}^{T^{-1}}$ and $R_{0}^{T}$, that is,\begin{equation}
(\Phi_{0}^{\prime}\Phi_{1}^{\prime-1})^{T}=B^{T^{-1}}(R_{0}R_{1}^{-1})^{T}B^{T}\label{eq:FiSimilarRTransposed}\end{equation}
We notice also that matrices $(R_{0}R_{1}^{-1})^{T}$ and $(\Phi_{0}^{\prime}\Phi_{1}^{\prime-1})^{T}$
are similar iff the matrices $R_{1}^{-1}R_{0}$ and $\Phi_{1}^{\prime-1}\Phi_{0}^{\prime}$
are also similar. Thus, the existence of a matrix $B$ such that the
matrices $(R_{0}R_{1}^{-1})^{T}$ and $(\Phi_{0}^{\prime}\Phi_{1}^{\prime-1})^{T}$
are similar is equivalent to the existence of a matrix $A$ such that
the matrices $R_{1}^{-1}R_{0}$ and $\Phi_{1}^{\prime-1}\Phi_{0}^{\prime}$
are similar. The existence of either of the matrices $A$ or $B$
in the eq. (\ref{eq:FiSimilarR}) or (\ref{eq:FiSimilarRTransposed})
is thus equivalent to the initial supposition that the states $|\psi\rangle$
and $|\psi^{\prime}\rangle$ are in the same SLOCC class.

We have therefore shown that \emph{two entangled states of dimensionality
$(n,n,2)$, $|\psi\rangle$ and $|\psi^{\prime}\rangle$ are interconvertible
through SLOCC iff, for any base $\{|r_{k}\rangle\}$ for the local
support of $|\psi\rangle$ in $C_{a}^{n}\otimes C_{b}^{n}$, a base
$\{|\phi_{k}^{\prime}\rangle\}$ for the local support of $|\psi^{\prime}\rangle$
in $C_{a}^{n}\otimes C_{b}^{n}$ exists such that the respective matrices
$R_{1}^{-1}R_{0}$ and $\Phi_{1}^{\prime-1}\Phi_{0}^{\prime}$ are
similar.}

From this result and Theorem 1, it is clear that two given states,
$|\psi\rangle$ and $|\psi^{\prime}\rangle$, are interconvertible
through some SLOCC only if they are in the same Jordan family. However,
if $|\psi\rangle$ and $|\psi^{\prime}\rangle$ belong to the same
Jordan family the situation is not to simple and we need further work
on eq. (\ref{eq:RelationEigenvalues}) to verify whether $|\psi\rangle$
and $|\psi^{\prime}\rangle$ are in the same SLOCC class. This can
be stated as follows. Let $|\psi\rangle$ and $|\psi^{\prime}\rangle$
be two entangled states of dimensionality $(n,n,2)$ and suppose we
get the bases $\{|r_{k}\rangle\}$ and $\{|r_{k}^{\prime}\rangle\}$
for the local support planes $\mathcal{P}$ and $\mathcal{P}^{\prime}$
in $C_{a}^{n}\otimes C_{b}^{n}$ from the states $|\psi\rangle$ and
$|\psi^{\prime}\rangle$ as in eq. (\ref{eq:plano}). Using the method
of section \ref{sub:Estados-Emaranhados-em-um-qubit}, we find that
the matrices $R_{1}^{-1}R_{0}$ and $R_{1}^{\prime-1}R_{0}^{\prime}$
are in the same Jordan family and have the eigenvalues $\{\lambda_{l,r}\}$
and $\{\lambda_{l,r^{\prime}}\}$ respectively, where the index  \textbf{$l=1,2,...L$}
($L$ being the number of distinct eigenvalues) is such that $\textrm{rank}(R_{1}^{-1}R_{0}-\lambda_{l,r})^{k}=\textrm{rank}(R_{1}^{\prime-1}R_{0}^{\prime}-\lambda_{l,r^{\prime}})^{k}$,
that is, the Jordan blocks corresponding to the eigenvalues $\lambda_{l,r}$
and $\lambda_{l,r^{\prime}}$ have the same structure. Thus we need
to verify whether we can find a base $\{|\phi_{k}^{\prime}\rangle\}$
for $\mathcal{P}^{\prime}$ such that the eigenvalues of the respective
matrix $\Phi_{1}^{\prime-1}\Phi_{0}^{\prime}$, $\{\mu_{l,\phi^{\prime}}\}$,
are all equal to $\lambda_{l,r}$ for each $l$. Using eq. (\ref{eq:RelationEigenvalues})
to get $\mu_{l,\phi^{\prime}}$ as a function of $\lambda_{l,r^{\prime}}$,
we must require that equation\[
\lambda_{l,r}=\mu_{l,\phi^{\prime}}=\frac{a\lambda_{l,r^{\prime}}+b}{c\lambda_{l,r^{\prime}}+d}\]
or\begin{equation}
\lambda_{l,r}\lambda_{l,r^{\prime}}c+\lambda_{l,r}d-\lambda_{l,r^{\prime}}a-b=0\label{eq:linearsystem}\end{equation}
to have at least one solution for all $l$'s with the additional condition
that $(ad-bc)=1$, that is, we have a linear system with $L$ equations
with an additional constraint for the variables $a$, $b$, $c$ and
$d$ which correspond to the coefficients of the linear transformation
$\Phi_{0}^{\prime}=aR_{0}^{\prime}+bR_{1}^{\prime}$ and $\Phi_{1}^{\prime}=cR_{0}^{\prime}+dR_{1}^{\prime}$.
Notice that $c\lambda_{l,r^{\prime}}+d\neq0$, since it is an eigenvalue
of $\Phi_{1}^{\prime-1}$, which is invertible. We observe that any
non-trivial solution of the linear system (\ref{eq:linearsystem})
intersects the surface defined by the additional constraint $(ad-bc)=1$
at two opposite points %
\footnote{In fact, this is true for any non-vanish value chosen for the determinant
$(ad-bc)$. This ultimately allows for making it equal to one. Moreover,
the resulting matrix $\Phi_{1}^{\prime-1}\Phi_{0}^{\prime}$ does
not depend on the value of this determinant.%
}, that is, if the linear system (\ref{eq:linearsystem}) has a non-trivial
solution, then there are always at least two solutions satisfying
also the additional constraint $(ad-bc)=1$ and differing by a sign.
In this way, we have reduced the problem of deciding if $|\psi\rangle$
and $|\psi^{\prime}\rangle$ are in the same SLOCC class to the existence
of a non-trivial solution of the homogeneous linear system (\ref{eq:linearsystem})
with $L$ equations.

In case the determinant of the linear system (\ref{eq:linearsystem})
has some nonvanishing minor of dimension larger than three, its unique
solution is the trivial one which is incompatible with the condition
$(ad-bc)=1$. Thus the states $|\psi\rangle$ and $|\psi^{\prime}\rangle$
will not be in the same SLOCC class. Hence, a Jordan family of states
with more than three distinct eigenvalues can be subdivided into an
infinity of SLOCC classes defined by the constraints that all minors
of (\ref{eq:linearsystem}) of dimension greater than three must vanish.
In case that the greatest nonvanishing minor of (\ref{eq:linearsystem})
has dimension three, then there are always two solutions which differ
only by a sign. Thus, in this case, we can find $a$, $b$, $c$ and
$d$ such that the transformation from the base $\{|r_{0}^{\prime}\rangle,|r_{1}^{\prime}\rangle\}$
to $\{|\phi_{0}^{\prime}\rangle,|\phi_{1}^{\prime}\rangle\}$ will
result in a matrix $\Phi_{1}^{\prime-1}\Phi_{0}^{\prime}$ similar
to $R_{1}^{-1}R_{0}$ and we can find also invertible local operators
$A$, $B$ and $C$ such that the expression (\ref{eq:LO-SLOCC})
holds, and so the states $|\psi\rangle$ and $|\psi^{\prime}\rangle$
are interconvertible through SLOCC. The operator $A$ can be obtained
from eq. (\ref{eq:FiSimilarR}). Similarly, the operator $B$ can
be obtained from eq. (\ref{eq:FiSimilarRTransposed}). The operator
$C$ can now be obtained from (\ref{PsiLinhaSLOCCPsi}). Hence, if
we have $L\leq3$ it is always possible to find a non-trivial solution
of (\ref{eq:linearsystem}), and every Jordan family with less than
three eigenvalues is equivalent to a SLOCC class.

When the greatest non-vanishing minor of (\ref{eq:linearsystem})
has dimension smaller than three, there will be one or two free parameters
in (\ref{eq:linearsystem}). These free parameters in principle may
allow for the existence of an infinity of SLOCC protocols depending
on SLOCC class. This may easily seen to be actually the case in specific
examples, e.g. two states in class W. However, examples also can be
found in which the matrix $\Phi_{1}^{\prime-1}\Phi_{0}^{\prime}$
turns out to be independent of the remaining free parameter, e.g.
two states in class GHZ.

When more than one distinct eigenvalue is associated with the same
Jordan block structure there will be more than one way to label the
two set of eigenvalues, $\{\lambda_{l,r}\}$ and $\{\lambda_{l,r^{\prime}}\}$.
The considered states, $|\psi\rangle$ and $|\psi^{\prime}\rangle$,
will be in the same SLOCC class provided at least one labelling can
be found for which we can get a solution to (\ref{eq:linearsystem}).
In many cases, it will be possible to find a solution to (\ref{eq:linearsystem})
for many labellings. Particularly, for two states in class GHZ, there
will be always a solution for each one of the two possible labellings.

\section{\label{sub:Casos-mais-gerais}Discussion: More General Tripartite
Entangled States}

It is easy to obtain an equation similar to (\ref{eq:superposicao})
for general tripartite systems. However, it appears to be very difficult
to classify the possible solutions. In the case involving one qubit
we have the bonus that we could transform the equation (\ref{eq:superposicao})
into an eigenvalue problem. In the general case it appears to be very
difficult to avoid being led to a system of polynomial equations in
many variables.

As an example, suppose we have an entangled state $|\psi\rangle$
of dimensionality $(n,n,n)$. Using the reasoning of section \ref{sub:Estados-Emaranhados-em-um-qubit},
we get that the sum in equation (\ref{eq:inicial}) now has $n$ terms
and the local support for the subsystem $s_{ab}$ is some $n$-dimensional
hyperplane $\mathcal{P}\in C_{a}^{n}\otimes C_{b}^{n}$. We could
easily get a base for this hyperplane using some bipartite decomposition
as done in (\ref{eq:inicial}) in section \ref{sub:Estados-Emaranhados-em-um-qubit},
i. e., we can write any state $|\phi\rangle\in\mathcal{P}$ as\[
|\phi\rangle=\sum_{k}\alpha_{k}|r_{k}\rangle,\]
where \textbf{$\alpha_{k}$} are complex coefficients and $|r_{k}\rangle\in C_{a}^{n}\otimes C_{b}^{n}$
is the relative state of $|k\rangle\in C_{c}^{n}$ , $\{|k\rangle\}$
being an orthonormal base in $C_{c}^{n}$. Using a base $\{|i\rangle\}$
in $C_{a}^{n}$ , a base $\{|j\rangle\}$ in $C_{b}^{n}$ and a state
\textbf{$|u_{a}\rangle\in C_{a}^{n}$}, we get an equation similar
to (\ref{eq:superposicao}), but with $n$ matrices\begin{equation}
(\sum_{k}\alpha_{k}R_{k})u_{a}^{*}=0\label{eq:MatrizGeral}\end{equation}
where, like in (\ref{eq:superposicao}), $R_{k}$ has the components
$[R_{k}]_{ij}=\langle ji|r_{k}\rangle$ and the vector $u_{a}^{*}$
has the components $u_{a_{i}}^{*}=\langle u_{a}|i\rangle$. Keeping
in mind that we are looking for the superpositions of matrices with
rank less than $n$, we can easily get from\begin{equation}
\det[\sum_{k}\alpha_{k}R_{k}]=0\label{eq:Determinate}\end{equation}
that there is at least one $(n-1)$-dimensional surface $S\subset\mathcal{P}$
in which the states have Schmidt rank $(n-1)$ or less. As the constraint
defined by equation (\ref{eq:Determinate}) is obviously non-linear,
there are $n$ states in $S$ that span $\mathcal{P}$. This mean
that there are infinitely many sub-Schmidt decompositions of $|\psi\rangle$
with $n(n-1)$ products. In order to identify the smaller sub-Schmidt
decompositions, we need to verify the existence of some set of $\alpha_{k}$'s
which makes null all $(n-1)$ minors of the matrix $(\sum_{k}\alpha_{k}R_{k})$.
For the simplest case with $n=3$, this gives us a system of nine
polynomials in two variables. Therefore, for general systems, we cannot
to avoid very unpractical conditions. However, we know that states
with sub-Schmidt decompositions exist, since we can explicitly write
entangled states of dimensionality $(n,n,n)$ with less than $n(n-1)$
products. As a example, we take some kind of general GHZ state in
$n$ dimensions,\begin{equation}
|\psi_{GHZ}\rangle=\sum_{k}d_{k}|a_{k}\rangle|b_{k}\rangle|c_{k}\rangle,\label{eq:generalGHZ}\end{equation}
where the $d_{k}$ are some complex coefficient and $\{|a_{k}\rangle\}$,
$\{|b_{k}\rangle\}$ and $\{|c_{k}\rangle\}$ are linearly independent
states in $C_{a}^{n}$, $C_{b}^{n}$ and $C_{c}^{n}$ respectively.
Particularly, it is easy to see that the family of all states that
can be written in the form (\ref{eq:generalGHZ}) are in a single
SLOCC class, since there are always invertible local operators $A$
in $n$ dimensions taking an arbitrary set of $n$ linearly independent
vectors $d_{k}|a_{k}\rangle$ into any other $n$ linearly independent
vectors $d_{k}^{\prime}|a_{k}^{\prime}\rangle$. Similarly for $B$
taking $|b_{k}\rangle$ into $|b_{k}^{\prime}\rangle$ and $C$ taking
$|c_{k}\rangle$ into $|c_{k}^{\prime}\rangle$. Thus there are always
invertible local operators $A$, $B$ and $C$ to satisfy (\ref{eq:LO-SLOCC}).

We must also observe that we cannot claim to have obtained a full
classification of entangled states in a space $C_{a}^{n}\otimes C_{b}^{n}\otimes C_{c}^{2}$.
In the space $C_{a}^{3}\otimes C_{b}^{3}\otimes C_{c}^{2}$ of example
2, for example, we can have factorable states, bipartite states of
the three types and the two classes of example 1, which were all previously
known, and furthermore the five additional classes of example 2. However,
in this space, we also have entangled states of dimensionality $(3,2,2)$.
In this case, when we use the method of section \ref{sub:Estados-Emaranhados-em-um-qubit}
we will get a local support in $C_{a}^{3}\otimes C_{b}^{3}$ with
no state with Schmidt rank 3 and almost all with Schmidt rank 2. Thus
all the matrices in $\mathcal{P}$ (\ref{eq:superposicao}) are non-invertible
and we cannot reduce the problem to an eigenvalue form like (\ref{eq:DeAutoValores}).
This case was in fact solved by Miyake and Verstraete \cite{Miyake2},
however there are many other cases in which the local supports of
Alice and Bob are distint and we do not know a solution, e. g., an
entangled state of dimensionality $(4,3,2)$.

\section{Conclusion}

We have described a constructive method to find decompositions of
tripartite entangled pure states which involve a number of terms smaller
than one obtains using two successive Schmidt decompositions. These
decompositions have been called sub-Schmidt decompositions for short.
Particularly for entangled states of dimensionality $(n,n,2)$, we
found a one-to-one correspondence between the concept of Jordan families
and the sub-Schmidt decompositions and use this correspondence to
classify all sub-Schmidt decompositions of entangled states in this
dimensionality. Moreover, from this classification of sub-Schmidt
decompositions, we got a classification of these states according
with their interconvertibility under SLOCC. We also briefly discussed
the difficulties in generalizing our methods to more general systems.
We expect that these results will contribute to the understanding
of higher dimensional and multipartite entanglement.

\begin{acknowledgments}
We would like to thank D. Tausk for crucial help on the proof of Theorem
1. MFC acknowledges financial support of FAPESP (Fundação de Amparo
a Pesquisa do Estado de São Paulo).
\end{acknowledgments}

\end{document}